\documentclass[a4paper]{spie}  
 \usepackage[centertags]{amsmath}
 \usepackage{amsfonts}
 \usepackage{amssymb}
 \usepackage[]{graphicx}
 \usepackage{amsmath}

\newcommand{\Neff}{\ensuremath{N_\mathit{eff}}}

\newcommand{\vc}{\mathbf{r}}

\title{Simple nonstationary generalization of Gaussian Shell-model
beams}

\author{Mikalai U. Karelin 
\skiplinehalf
 B. I. Stepanov Institute of Physics, National Academy of
 Sciences \\
 Minsk, 220072 Belarus} 

\authorinfo{Further author information:\\ 
 E-mail: karelin@dragon.bas-net.by, Telephone: (+375 17) 284 1419, Fax: 284 0879}

\begin{document}
\maketitle


\begin{abstract}
Model of partially coherent pulses, based on the concept
of ``hidden coherence'', introduced recently by Picozzi and
Haelterman in framework of parametric wave mixing, is presented.
The nonuiform and nonstationary phase shift, while completely
deterministic, results in the beam properties, which are typical
for partially coherent light --- low contrast of interference
effects, increase of spectral width and so on --- i.~e. light becomes
effectively non-coherent. The proposed model is studied in
framework of coherent mode decomposition, its main properties and
limitations of the model are discussed.
\end{abstract}

\keywords{Coherence, pulsed beam, phase shift, Shell-model}

\section{Introduction and motivation}
\label{sect:intro}

Nowadays there is a constant interest to use the partially
coherent light in optical systems. Among the important problems 
related to partially coherent light are study of light propagation,
image formation, pulse shaping, light interaction with nonlinear 
media, etc. All of these problems require adequate models of radiation,
which should describe its spatial, spectral and statistical
properties; at the same time analytically described models are 
advantageous. For stationary beams such model is well-known: it is
the Shell model \cite{MandelWolf1995}, especially Gaussian 
Shell-model (GSM) and its generalizations. Pulsed partially coherent 
beams is also described by various generalizations of the Shell-model 
\cite{Paakkonen_etal2002,LiWang_etal2003,Christov1985,Christov1986,Aleshkevich_etal1988}, 
usually with the Gaussian profile of temporal dependence of 
correlation function.

In the present work, another model of partially coherent pulses is
presented. It is based on the concept of ``hidden coherence'',
introduced recently by Picozzi and Haelterman 
\cite{PicozziHaelterman2002} in the framework of
parametric wave mixing. Such radiation is completely coherent along 
some spatio-temporal trajectories in 4D space, but is neither completely 
spatially nor temporally coherent (from the point of view of usual 
interferometric experiments, such as Young double-slit interferometer).
In the other words, wavefields with complex spatio-temporal structure are 
``operationally'' equivalent to partially coherent beams 
\cite{LazarukKarelin1997a}.

As an example, nonuiform and nonstationary phase shift (completely deterministic, 
such as in non-linear self-phase modulation \cite{Ivakin_etal2002}, 
or, for relatively large characteristic time intervals --- by 
mechanical motion of lenses, mirrors, or phase screens
\cite{Mendlovic_etal1999,Lohmann_etal1999b}) results in the beam
properties, which are typical for partially coherent light --- low
contrast of interference effects, increase of spectral width and
so on --- i.e. light becomes effectively non-coherent.

In present paper, a quite a simple model of effectively partially coherent 
pulses is studied in framework of coherent mode decomposition, its 
main properties and limitations of the model are discussed.

\section{Modal approach to partial coherence description 
(Karhunen-Lo\'eve decomposition)}

A complex field (analytic signal) of a light beam cross-section is
considered as sum of factorized, mutually orthogonal (and
normalized) components
 $$
 E(\vc,t) = \sum\nolimits_k \nu_k \;\mathcal{E}_k(\vc)\,e_k(t),
 $$
$\nu_k$ is amplitude of $k$th component.


\begin{itemize}
	\item Envelopes and modal amplitudes are calculated as 
eigenfunctions and eigenvalues of dual integral equations where 
kernels are spatial and temporal correlation functions
 $$
 |\nu_k|^2 \mathcal{E}_k(\vc) = \int d\vc \, \Gamma_S(\vc,\vc') \, 
    \mathcal{E}^{\ast}(\vc'),
 $$
 $$
 |\nu_k|^2 e_k(t) = \int dt \, \Gamma_T(t,t') \, e^{\ast}(t').
 $$
  \item Decomposition gives optimal representation (most fast 
converging expansion).
	\item Effective number of modes
 $$
 \Neff = \sum\nolimits_k |\nu_k|^4 \big/ 
               \left(\sum\nolimits_k |\nu_k|^2 \right),
 $$	
 $$
 \Neff = \frac{\left(\int_{-\infty}^\infty dx
                        \, \Gamma_S(x,x)\right)^2}
   {\int_{-\infty}^\infty dx \int_{-\infty}^\infty dx'
         \, |\Gamma_S(x,x')|^2},
       = \frac{\left(\int_{-\infty}^\infty dt
                        \, \Gamma_T(t,t)\right)^2}
   {\int_{-\infty}^\infty dt \int_{-\infty}^\infty dt'
         \, |\Gamma_T(t,t')|^2}
 $$
(or overall degree of coherence $\mu = 1/\Neff$) characterize 
contrast of interference effects (e.g. speckles)
 $$
 C = \Neff^{-1/2}
 $$
\end{itemize}

\section{1D Model}

Light pulse with the spatially nonuiform and nonstationary phase 
shift:
 $$
 E(x,t) = \frac{A}{\sqrt[4]{4\pi^2 a^2 \tau^2}}
 \exp\left(- \frac{x^2}{4 a^2} - \frac{t^2}{4 \tau^2}
 + i t\,x/\eta \right)
 \exp(-i \omega_0 t),
 $$
$\eta$ --- phase shift parameter.

Its spatial and temporal correlation functions
 $$
 \Gamma_S(x,x') = \int_{-\infty}^\infty dt \, E(x,t) E^\ast(x',t),
 $$
 $$
 \Gamma_T(t,t') = \int_{-\infty}^\infty dx \, E(x,t) E^\ast(x,t')
 $$
are of GSM form in both domains
 $$
 \Gamma_S(x,x') = \frac{A^2}{\sqrt{2\pi a^2}}
 \exp\left(- \frac{x^2+x'^2}{4 a^2} - \frac{(x-x')^2}{2 \sigma_S^2}\right),
 $$
 $$
 \Gamma_T(t,t') = \frac{A^2}{\sqrt{\pi \tau^2}}
 \exp\left(- \frac{t^2+t'^2}{4 \tau^2} - \frac{(t-t')^2}{2 \sigma_T^2} \right)
 $$
with $\sigma_S = \eta/\tau$, $\sigma_T = \eta/a$.

Note, that an observer, moving with a constant velocity 
$\propto 1/\eta$ across the beam, will treat it as fully coherent.

\section{The Model Properties}

\begin{itemize}
	\item Fully deterministic.
	\item Spatio-temporal symmetry.
	\item Any degree of coherence.
	\item Could be used in quasi-1D (or $1+1+1$D) problems, e.~g. with 
strip-source illumination.
\end{itemize}

\textbf{Disadvantage:} the proposed model requires large phase shifts 
at beam edges and on pulse rise and fall.

\textbf{Generation:} using specially designed moving DOE (diffractive 
optical element). Possibly together with lens and a (chirp-like) 
phase modulator.

\textbf{Approximation:} Nonlinear self-phase modulation 
 $$
 E_\mathrm{out}(x,t) = E_\mathrm{in}(x,t) 
     \exp\left[i\alpha/\eta |E_\mathrm{in}(x,t)|^2 \right]
 $$
($\alpha \approx 3$) converts initially coherent pulsed beam 
$E_\mathrm{in}(x,t) = \mathcal{E}(x) e(t)$ into light with 
nonstationary and nonuniform phase shift and with approximately 
Gaussian correlation function (error is within $5\%$).

\section{Coherent mode decomposition}

Spatial and temporal decomposition functions are already known --- 
Hermite-Gaussian functions
 $$
 \mathcal{E}_k(x) = \left(\frac{2c}{\pi}\right)^{1/4} 
        \frac{1}{(2^k \, k!)^{1/2}}
        H_k(x \, (2c)^{1/2}) \, \exp(-c x^2)
 $$
 $$
 e_k(t) = \left(\frac{2d}{\pi}\right)^{1/4} 
        \frac{1}{(2^k \, k!)^{1/2}}
        H_k(t \, (2d)^{1/2}) \, \exp(-d t^2)
 $$
and modal weights
 $$
 |\nu|^2 = A^2 \frac{\eta^2+b}{\eta^2+2a^2\tau^2+b}
    \left(\frac{2a^2\tau^2}{\eta^2+2a^2\tau^2+b}\right)^n
 $$
where $H_k(x)$ are Hermitian polynomials, $b = \sqrt{\eta^2 + 
4 a^2\tau^2}$ and
 $$
 c = \frac{\eta^2 + 4 a^2\tau^2}{16 a^4 \eta^2}, 
 \quad 
 d = \frac{\eta^2 + 4 a^2\tau^2}{16 \tau^4 \eta^2}.
 $$
 
\bigskip 
Number of coherent modes 
 $$
 \Neff \propto a \tau/\eta, \quad \text{for} \; \eta \ll 1.
 $$

\section{Discussion of 2D case}

One more disadvantage of the model: difficulty of its
generalization to more realistic case of 2D aperture. 
It is preferable, that field is Rotationally invariant, then 
use of term like 
 $$
 E(x,y,t) \propto \exp\left(i \,\eta\,\sqrt{x^2 + y^2}\,t\right)
 $$
leafs to a phase singularity near $x = y = 0$. An alternative variant 
with 
 $$
 E(x,y,t) \propto \exp\left(i \,\eta\,(x^2 + y^2)\,t\right)
 $$
does not lead to Gaussian shape of spatial correlation function.

Moreover, in general, taking into account properties of modal 
decomposition of 1D (temporal $\Gamma_T(t,t')$) and 2D spatial 
$\Gamma_S(x,x')$ kernels, it is possible to show, that 
\textit{deterministic} double-Gaussian Shell-model pulsed beams are 
impossible.

\section*{Acknoledgement}

The work has been supported by Belarusian Fund for Fundamental 
Research, projects No. F03MS-066 and F05K-056.

 
\end{document}